\documentclass[preprint,12pt]{elsarticle}

\usepackage{amssymb}
\usepackage{amsthm}
\usepackage{amsmath}
\usepackage{bm}

\begin{document}

\begin{frontmatter}



\title{Late-time cosmology in a model of modified gravity with an exponential function of the curvature}


\author[1]{A.~Oliveros}
\ead{alexanderoliveros@mail.uniatlantico.edu.co}

\author[1]{Mario A.~Acero}
\ead{marioacero@mail.uniatlantico.edu.co}

\affiliation[1]{organization={Programa de Física, Universidad del Atlántico},
    addressline={Carrera 30 Número 8-49}, 
    city={Puerto Colombia},
    state={Atlántico},
    country={Colombia}}


\begin{abstract}
In this work, we analyse the late-time evolution of the Universe for a particular $f(R)$ gravity model built from an exponential function of the scalar curvature. Following the literature, we write the field equations in terms of a suited statefinder function ($y_H(z)$) and considering well motivated physical initial conditions, the resulting equations are solved numerically. Also, the cosmological parameters $w_{\rm{DE}}$, $w_{\rm{eff}}$, $\Omega_{\rm{DE}}$ and $H(z)$ and the statefinder quantities $q$, $j$, $s$ and $Om(z)$ are explicitly expressed in terms of $y_H(z)$ and its derivatives. Furthermore, setting an appropriate set of values for the model parameters, the cosmological parameters as well as the statefinder quantities are plotted, and their present values (at $z=0$), are shown to be compatible with Planck 2018 observations and the $\Lambda$CDM--model values. Considering  updated measurements from the dynamics of the expansion of the Universe, $H(z)$, we perform an statistical analysis to constrain the free parameters of the model, finding a particular set of values that fit the data well and predict acceptable values for the cosmological and statefinder parameters at present time. Therefore, the $f(R)$ gravity model is found to be consistent with the considered observational data, and a viable alternative to explain the late-time acceleration of the Universe.
\end{abstract}


\end{frontmatter}


\section{Introduction}\label{sec_intro}
Modified gravity theories are a set of proposals used in last years in order to explain the observed late-time acceleration of the Universe, among other cosmological and astrophysical phenomena. In these scenarios, it is not necessary to consider dark energy (DE) or new forms of matter to explain the late-time acceleration (for more details about this topic see Refs.~\citep{peebles,copeland,bamba1}). 
Modified gravity theories include scalar-tensor theories, scalar-vector-tensor theories, Einstein–aether, Bimetric theories, TeVeS, $f(R)$, general higher-order theories, Ho\v{r}ava–Lifschitz gravity, Galileons, Ghost Condensates, and models of extra dimensions, including Kaluza-Klein, Randall–Sundrum and DGP (see e.g.~Refs. \citep{odintsov1,clifton,odintsov2} for a review). Among this wide spectrum of proposals, $f(R)$ gravity is one of the most studied in the literature, \citep{hwang, nojiri1,
capozziello1, nojiri2, capozziello2, nojiri3, hu1, mao, olmo, hu2, faraoni, bean, nojiri4, capozziello3, sasaki, appleby, carloni, capozziello4, elizalde, capozziello5, odintsov3, odintsov4, odintsov5, oikonomou1, oikonomou2} since in this framework, it is possible to analyse and unify different cosmological eras (inflation and dark energy). In general, the explicit form of the function $f(R)$ is given by some consistency requirements and diverse constraints, which impose conditions for the cosmological viability of $f(R)$ dark energy models \citep{amendola}, i.e., the function $f(R)$ must satisfy both cosmological and solar-system tests in the small-field limit of the parameter space. Particular $f(R)$ models satisfying such requirements that have been studied intensively in the literature include: the Hu-Sawicki model \citep{hu3}, Starobinsky's model \citep{starobinsky}, Tsujikawa's model \citep{tsujikawa}, and the Exponential model \citep{linder}. More recently, the Gogoi-Goswami model \citep{gogoi} has attracted special attention. Usually in these models, the particle associated with the cold dark matter component is not consider explicitly. In this sense, the authors of Ref.~\citep{odintsov5} have considered an axion $f(R)$ gravity model, in which the axion is the main component of cold dark matter in the Universe, demonstrating that the such a model can unify the early-time with the late-time acceleration (see Refs.~\citep{profumo,marsh,co,oikonomou4} for a more exhaustive exposition about this topic).

In this work, we carry out a detailed study of the late-time evolution of the Universe taking into account the $f(R)$ gravity model introduced in Ref.~\citep{granda}, which is based on an exponential function of the scalar curvature. This model behaves akin to $\Lambda$CDM at early times, and satisfies local and cosmological constraints. The procedure that we follow here, consists in writing the field equations in terms of a suitable statefinder function and, using well motivated  physical initial conditions, solving the resulting equations numerically. This strategy has been exhaustively implemented in the literature \citep{hu3, bamba, odintsov6, odintsov7, oikonomou3}, but until now, it has not been applied for the $f(R)$ gravity model under consideration. This study is mainly focused in analysing the behavior of some cosmological parameters ($w_{\rm{DE}}$, $w_{\rm{eff}}$, $\Omega_{\rm{DE}}$ and $H(z)$) together with the statefinder quantities $q$, $j$, $s$ and $Om(z)$. Thereby, with this analysis we hope to broaden the evidence which allows us to validate or to rule out this $f(R)$ gravity model. To add up to such scrutiny, we perform a statistical analysis, using  updated measurements
from the dynamics of the expansion of the Universe, $H(z)$, looking for establishing restrictions to the free parameter of the model.

This paper is organized as follows: in section \ref{sec_model} we perform a brief review about the $f(R)$ gravity. In section \ref{sec_cosmo}, we write the field equations in terms of a suitable statefinder function, $y_H(z)$, and taking into account physically motivated initial conditions, the resulting equations are solved numerically. In this section we also explicitly present the cosmological parameters $w_{\rm{DE}}$, $w_{\rm{eff}}$, $\Omega_{\rm{DE}}$, and $H(z)$, and the statefinder quantities $q$, $j$, $s$, and $Om(z)$, in terms of $y_H(z)$ and its derivatives. Furthermore, considering appropriate values for the model parameters, the evolution of the cosmological parameters and the statefinder quantities is plotted with respect to the redshift $z$, carrying out a complete cosmological analysis. Finally, in section \ref{sec_fit} we consider a set of observational data of $H(z)$, to constrain the distinct parameters of the model under consideration. Our conclusions are exposed in section \ref{conclusions}. 

\section{The $f(R)$ gravity: a brief review}\label{sec_model} 
In general, the action for an $f(R)$ gravity model in the presence  of matter components is given by
\begin{equation}\label{eq1}
S=\int{d^4x\sqrt{-g}\left(\frac{f(R)}{2\kappa^2}+\mathcal{L}_M\right)},
\end{equation}
where $g$ denotes the determinant of the metric tensor $g^{\mu\nu}$, $\kappa^2=8\pi G=1/M_p^2$, with $G$ being the Newton's constant and $M_p$ the reduced Planck mass. $\mathcal{L}_M$ represents the Lagrangian density for the matter components (relativistic and non-relativistic perfect matter
fluids). The term $f(R)$ is for now an arbitrary function of the Ricci scalar $R$. Variation with respect to the metric gives the equation
of motion
\begin{equation}\label{eq2}
f_R(R)R_{\mu\nu}-\frac{1}{2}g_{\mu\nu}f(R)+(g_{\mu\nu}\square-\nabla_\mu\nabla_\nu)f_R(R)=\kappa^2T_{\mu\nu}^{(M)},
\end{equation}
where $f_R\equiv \frac{df}{dR}$, $\nabla_{\mu}$ is the covariant derivative associated with the Levi-Civita connection of the metric, and
$\square\equiv \nabla^\mu\nabla_\mu$. Plus, $T_{\mu\nu}^{(M)}$ is the matter energy–momentum tensor which is assumed to be a perfect fluid.
Considering the flat Friedman-Robertson- Walker (FRW) metric,
\begin{equation}\label{eq3}
ds^2=-dt^2+a(t)^2\delta_{ij}dx^idx^j,
\end{equation}
with $a(t)$ representing the scale factor,  the time and spatial components of Eq.~(\ref{eq2}) are given, respectively, by
\begin{equation}\label{eq4}
3H^2f_R=\kappa^2\rho_M+\frac{1}{2}(Rf_R-f)-3H\dot{f}_R,
\end{equation}
and
\begin{equation}\label{eq5}
-2\dot{H}f_R=\kappa^2(\rho_M+P_M)+\ddot{f}_R-H\dot{f}_R,
\end{equation}
where $\rho_M$ and $P_M$ denote the density and pressure, respectively, of any perfect fluid of non-relativistic and relativistic matter. As usual in the literature, it is possible to rewrite the field equations (\ref{eq4}) and (\ref{eq5}) in the Einstein-Hilbert form:
\begin{equation}\label{eq6}
3H^2=\kappa^2\rho,
\end{equation}
\begin{equation}\label{eq7}
-2\dot{H}^2=\kappa^2(\rho+P),
\end{equation}
where $\rho=\rho_M+\rho_{\rm{DE}}$ and $P=P_M+P_{\rm{DE}}$ correspond to the total effective energy density and total effective pressure density of the cosmological fluid. Note that $\rho_M=\rho_m+\rho_r$. In this case, the dark energy component has a geometric origin, and after a some manipulation in Eqs.~(\ref{eq4})  and (\ref{eq5}), we obtain
\begin{equation}\label{eq8}
\rho_{\rm{DE}}=\frac{1}{\kappa^2}\left[\frac{Rf_R-f}{2}+3H^2(1-f_R)-3H\dot{f}_R\right],
\end{equation}
\begin{equation}\label{eq9}
P_{\rm{DE}}=\frac{1}{\kappa^2}[\ddot{f}_R-H\dot{f}_R+2\dot{H}(f_R-1)-\kappa^2\rho_{\rm{DE}}].
\end{equation}
Following the notation used in Ref.~\citep{odintsov6}, we shall assume that the perfect fluids, composed of Cold Dark Matter (CDM) and radiation, can be written as
\begin{equation}\label{eq10}
\rho_M(t)=\rho_{m0}\frac{1}{a^3}\left(1+\chi\frac{1}{a}\right),
\end{equation}
where $\rho_{m0}$ is the energy density of the CDM at present time, and $\chi=\rho_{r0}/\rho_{m0}=\Omega_{r0}/\Omega_{m0}\simeq 3.1\times 10^{-4}$, with $\rho_{r0}$ being the current density of relativistic component.  
\section{Late-time cosmological evolution}\label{sec_cosmo}
In this section we analyse the late-time evolution of the Universe taking into account a particular $f(R)$ gravity model.
Since the set of equations obtained here are very complicated, it will be necessary to solve them numerically. In order to facilitate such implementation, we perform a change of variable, i.e., we use the redshift, $z$, instead of the cosmic time $t$ as a dynamical variable.

To begin with, we know that
\begin{equation}\label{eq11}
1+z=\frac{1}{a(t)},
\end{equation}
where $a(0)=1$ is assumed. Then, from Eq.~(\ref{eq10}), it is easy to derive the following relation:
\begin{equation}\label{eq12}
\frac{d}{dt}=-H(1+z)\frac{d}{dz}.
\end{equation}
In this way, it is evident that
\begin{equation}\label{eq13}
\dot{H}=-H(z)(1+z)H'(z),
\end{equation}
where prime denotes derivative with respect to the redshift $z$; in a similar fashion, we can write $R$ and $\dot{f}_R$ as follows:
\begin{equation}\label{eq14}
R(z)=6(2H^2+\dot{H})=6[2H(z)^2-(1+z)H(z)H'(z)],
\end{equation}

\begin{equation}\label{eq15}
\begin{split}
\dot{f}_R(z) &= \dot{R}f_{RR} \\
             &= 6H(z)(1+z)^2\left[H'(z)^2+H(z)H''(z)
             -\frac{3H(z)H'(z)}{1+z}\right]f_{RR}. 
\end{split}
\end{equation}
It is clear in this context that we only need to use Eq.~(\ref{eq6}) for our analysis, but instead of solving the resulting equation for $H(z)$, we introduce a suitable statefinder function $y_H(z)$, defined by \citep{hu3, bamba, odintsov6, odintsov7, oikonomou3}
\begin{equation}\label{eq16}
y_H(z)\equiv \frac{\rho_{\rm{DE}}}{\rho_{m0}}=\frac{H(z)^2}{m_s^2}-(1+z)^3-\chi(1+z)^4,
\end{equation}    
where $m_s^2=\frac{\kappa^2\rho_{m0}}{3}=\Omega_{m0}H_0^2\simeq 6.51108\times 10^{-67}\,\rm{eV}^2$.\footnote{This value is obtained using the latest Planck data \citep{planck}: $\Omega_{m0}\simeq 0.315$ and $\rm{h}\simeq 0.674$, and taking into account the conversion factors: $1\,\rm{s}=1.51927\times 10^{15}\,\rm{eV}^{-1}$ and $1\,\rm{m}=5.06765\times 10^{6}\,\rm{eV}^{-1}$, which follows from considering the natural units, $\hbar = c = 1$. Hence, $H_0=100\,\rm{h}\,\rm{km}\,\rm{s}^{-1}\,\rm{Mpc}^{-1}=2.1331\, \rm{h}\times 10^{-42}\,\rm{GeV}=1.4377\times 10^{-33}\,\rm{eV}$.} From Eq.~(\ref{eq16}), we obtain the following useful expressions in terms of $y_H(z)$ and its derivatives: 

\begin{eqnarray}\label{eq17}
H(z)^2 &=& m_s^2[y_H(z)+(1+z)^3(1+(1+z)\chi)], \nonumber
\\
H(z)H'(z) &=& \frac{1}{2}m_s^2[y_H'(z)+(1+z)^2(3+4(1+z)\chi)],
\\
H'(z)^2+H(z)H''(z) &=& \frac{1}{2}m_s^2[y_H''(z)+6(1+z)(1+2(1+z)\chi)].\nonumber
\end{eqnarray}
Therefore, Eqs.~(\ref{eq14}) and (\ref{eq15}) take the following form:
\begin{equation}\label{eq20}
R(z)=3m_s^2[4y_H(z)-(1+z)y_H'(z)+(1+z)^3],
\end{equation}
\begin{equation}\label{eq21}
\begin{split}
\dot{f}_R(z) = &-3m_s^3(1+z)\sqrt{y_H(z)+(1+z)^3(1+(1+z)\chi)}\\
& \times[3y_H'(z)+(1+z)(3(1+z)-y_H''(z))]f_{RR},
\end{split}
\end{equation}
and  Eq.~(\ref{eq10}) in terms of the redshift is
\begin{equation}\label{eq22}
\rho_M(z)=\frac{3m_s^2}{\kappa^2}(1+z)^3[1+(1+z)\chi].
\end{equation}
Similarly, the cosmological parameters which will be used later, are given by:
\begin{equation}\label{eq23}
\Omega_{\rm{DE}}(z)=\frac{y_H(z)}{y_H(z)+(1+z)^3[1+\chi(1+z)]},
\end{equation}
\begin{equation}\label{eq24}
w_{\rm{DE}}(z)=-1+\frac{1}{3}(1+z)\frac{y'_H(z)}{y_H(z)},
\end{equation}                                     
\begin{equation}\label{eq25}
\begin{split}
w_{\rm{eff}}(z) &=-1+\frac{2}{3}(1+z)\frac{H'(z)}{H(z)} \\ 
&=\frac{(1+z)y'_H(z)-3y_H(z)+(1+z)^4\chi}{3\{y_H(z)+(1+z)^3[1+\chi(1+z)]\}}.
\end{split}
\end{equation}

The statefinder parameters are also rewritten in the following form:
\begin{equation}\label{eq26}
\begin{split}
q(z) &= -1+(1+z)\frac{H'(z)}{H(z)} \\
    & = \frac{(1+z)y'_H(z)-2y_H(z)+(1+z)^3[1+2\chi(1+z)]}{2\{y_H(z)+(1+z)^3[1+\chi(1+z)]\}},
\end{split}
\end{equation}
\begin{equation}\label{eq27}
\begin{split}
j(z)&=-2-3q+\frac{\ddot{H}}{H^3}\\
&=\frac{2y_H(z)+(1+z)\{-2y'_H(z)+(1+z)[2(1+z)(1+3(1+z)\chi)+y''_H(z)]\}}{2\{y_H(z)+(1+z)^3[1+\chi(1+z)]\}},
\end{split}
\end{equation}
\begin{equation}\label{eq28}
s(z)=\frac{j-1}{3(q-\frac{1}{2})}=\frac{(1+z)[(1+z)y''_H(z)-2y'_H(z)+4(1+z)^3\chi]}{3(1+z)[(1+z)^3\chi+y'_H(z)]-9y_H(z)},
\end{equation}
\begin{equation}\label{eq29}
\begin{split}
Om(z) &= \frac{\left(\frac{H}{H_0}\right)^2-1}{(1+z)^3-1} \\ 
      &=\frac{y_H(z)+(1+z)^3[1+(1+z)\chi]-1-\chi-y_H(0)}{[(1+z)^3-1][1+\chi+y_H(0)]},
\end{split}
\end{equation}
where the expressions given in Eq.~(\ref{eq17}) were used systematically. Note that these quantities depend explicitly only on the parameter $\chi$.

It is at this point that we introduce the $f(R)$ gravity model which plays the central role in this work:
\begin{equation}\label{eq30}
f(R)=R-2\,\lambda\,\mu^2\,e^{-(\mu^2/R)^\eta},
\end{equation}
where $\lambda$ and $\eta$ are positive real dimensionless parameters, and $\mu$ is a positive real parameter with dimension of $\rm{eV}$. 

This model was introduced in Ref.~\citep{granda}, and it behaves very close to $\Lambda$CDM at early times and satisfy local and cosmological constraints. The model inserts an exponential function of the scalar curvature, which works as a correction to the Einstein term and disappears in the limit $R \rightarrow 0$, containing flat space-time solutions and allowing the possibility of pure geometrical explanation of the dark energy phenomenon. An extension of this model (where an $R^2$ Starobinsky term is added) able to explain early time inflation and late time accelerated expansion was studied in Ref.~\citep{granda2}. Although such an extension, i.e., including an extra $R^2$ term in Eq.~(\ref{eq30}), might be of some interest (see \cite{odintsov10} for a recent discussion about this subject), we have gone through the numerical solution of the relevant equations to demonstrate that its contribution is negligible at late-times (dark energy era). Then, in  Ref.~\citep{granda3}, the author performs a generalization of this model, introducing a general function of the scalar curvature in the exponential term. In the literature, other authors have studied some $f(R)$ gravity models with exponential functions of the scalar curvature (see for example Refs.~\citep{linder} and \citep{odintsov8}).

\begin{figure*}
\centering
    \includegraphics[width=0.47\textwidth]{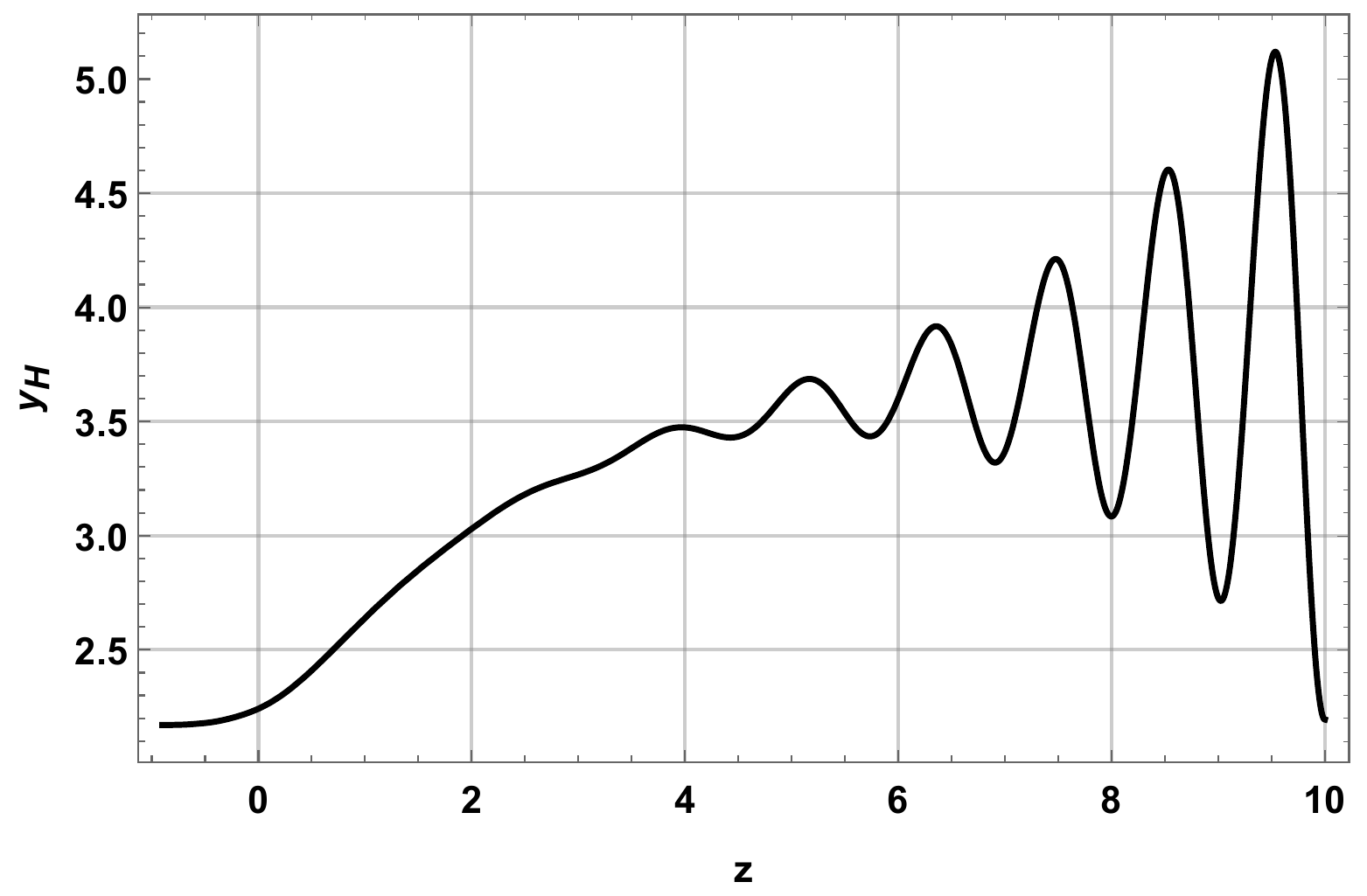}
    \includegraphics[width=0.47\textwidth]{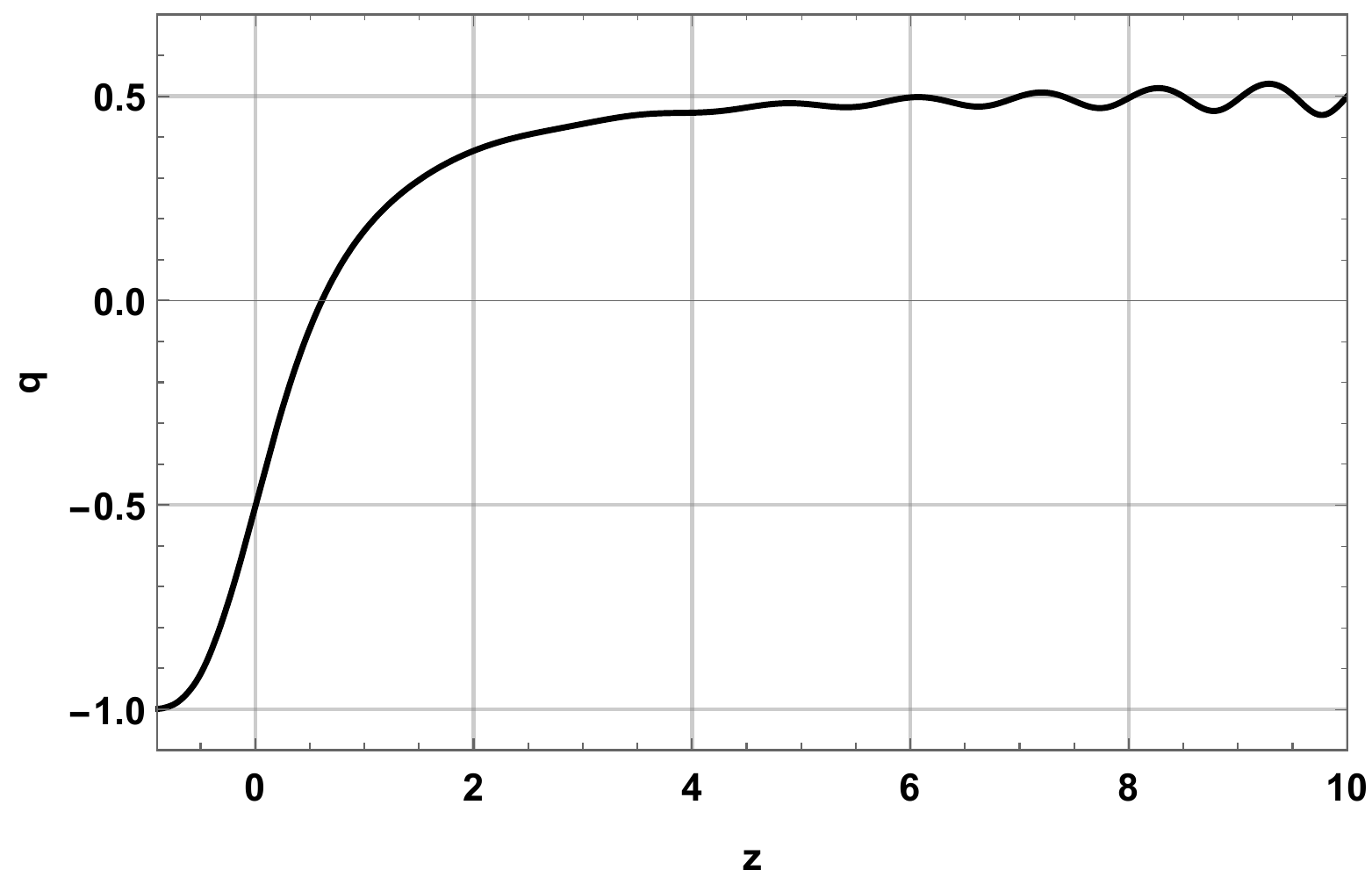}
    \caption{Evolution of the statefinder function $y_H(z)$ (\emph{left}) and the deceleration parameter $q(z)$ (\emph{right}), with the redshift, $z$.}
    \label{fig1}
\end{figure*}

\begin{figure*}
\centering
    \includegraphics[width=0.47\textwidth]{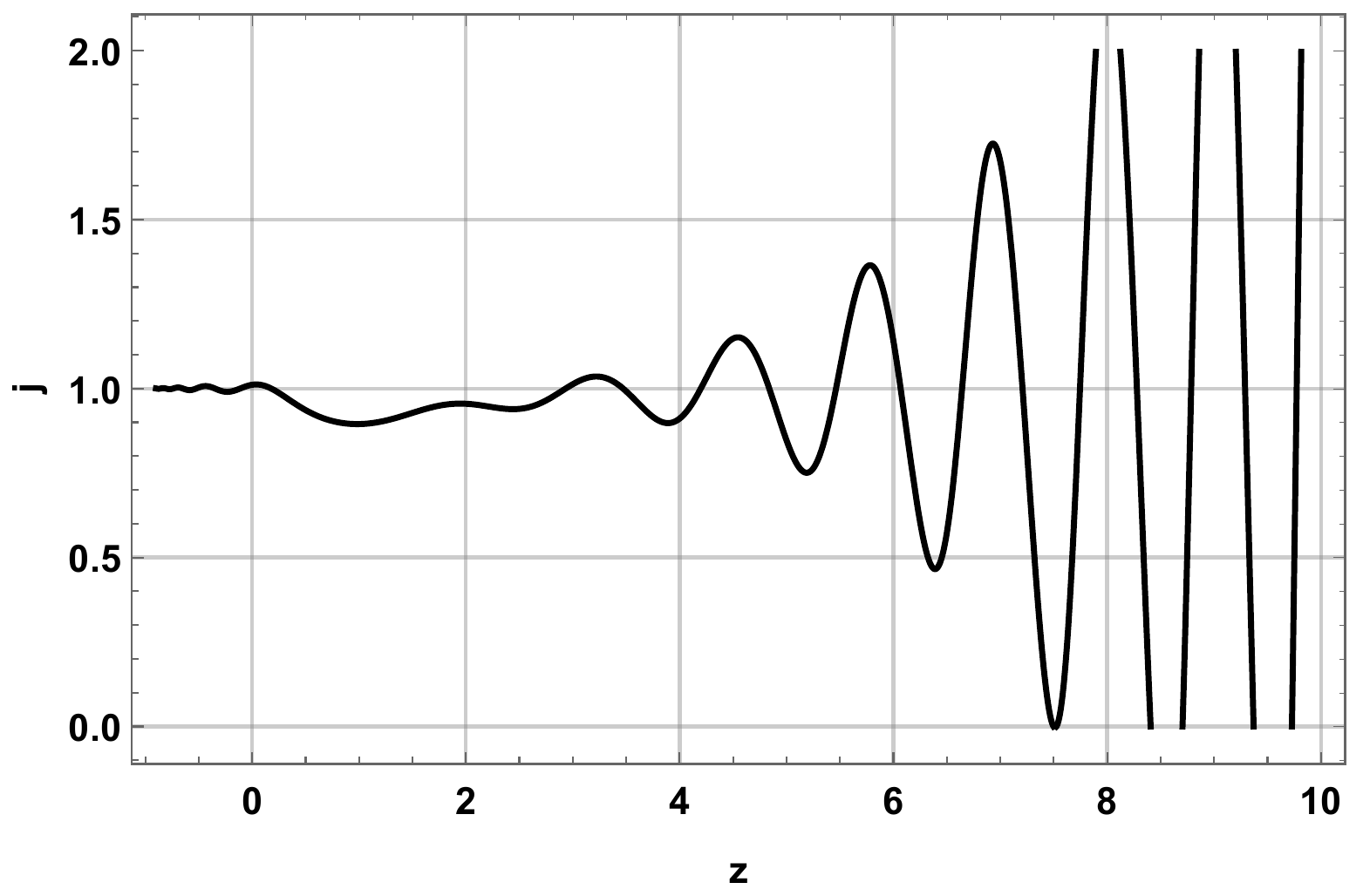}
    \includegraphics[width=0.47\textwidth]{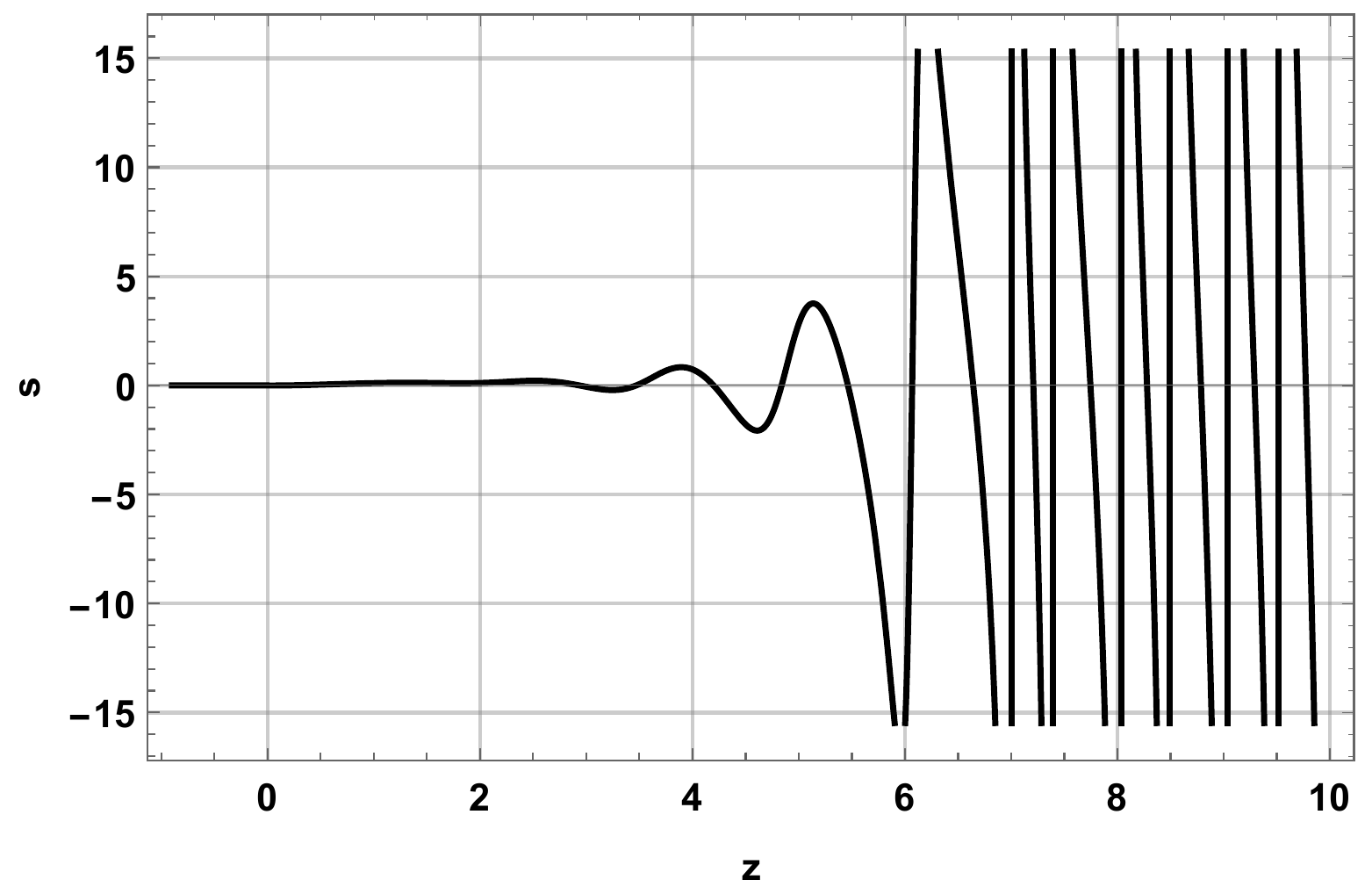}
    \caption{The statefinder parameters $j(z)$ (\emph{left}) and $s(z)$ (\emph{right}) as a function of the redshift, $z$.}
    \label{fig2}
\end{figure*}

\begin{figure*}
\centering
    \includegraphics[width=0.47\textwidth]{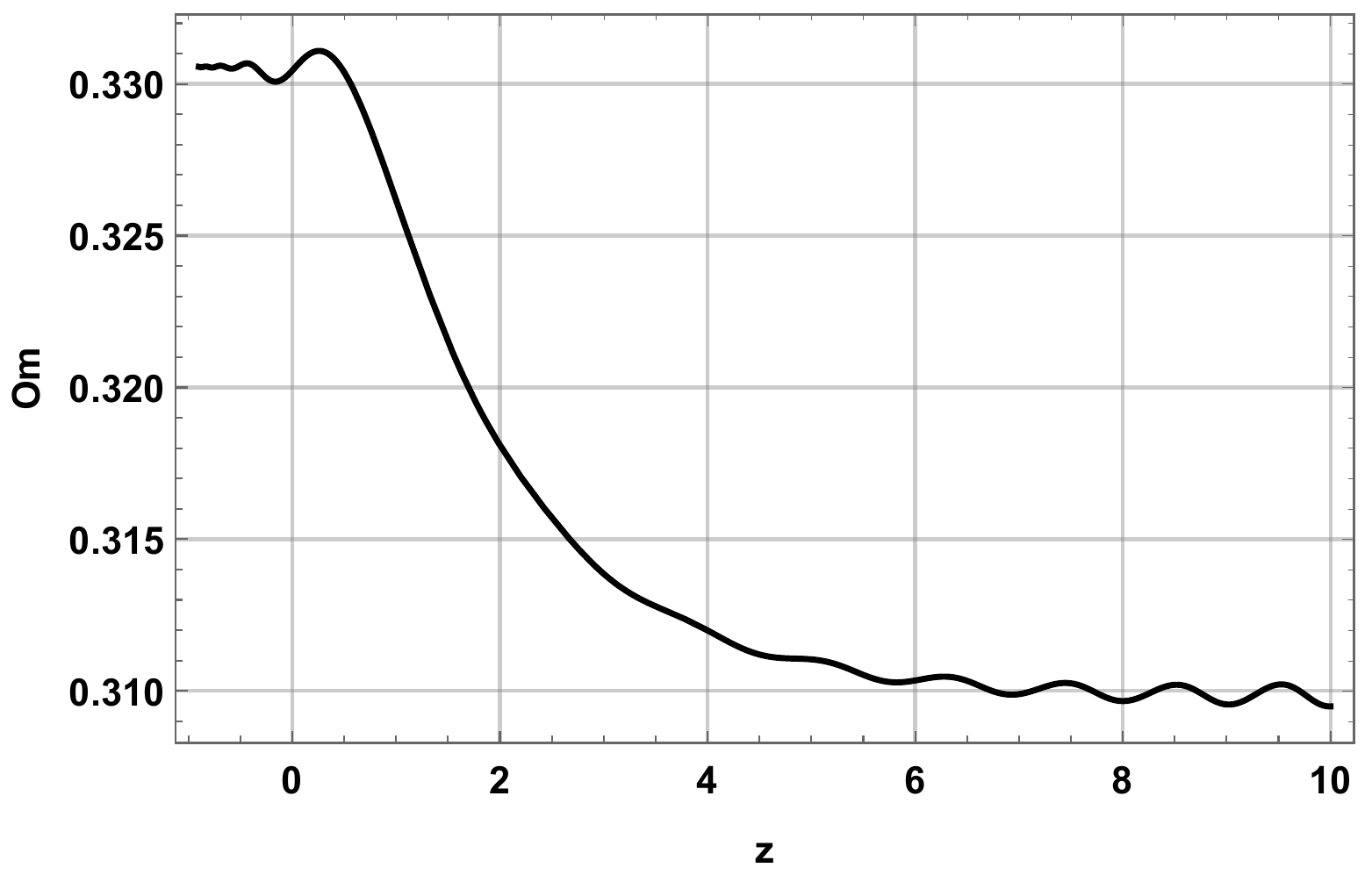}
    \includegraphics[width=0.47\textwidth]{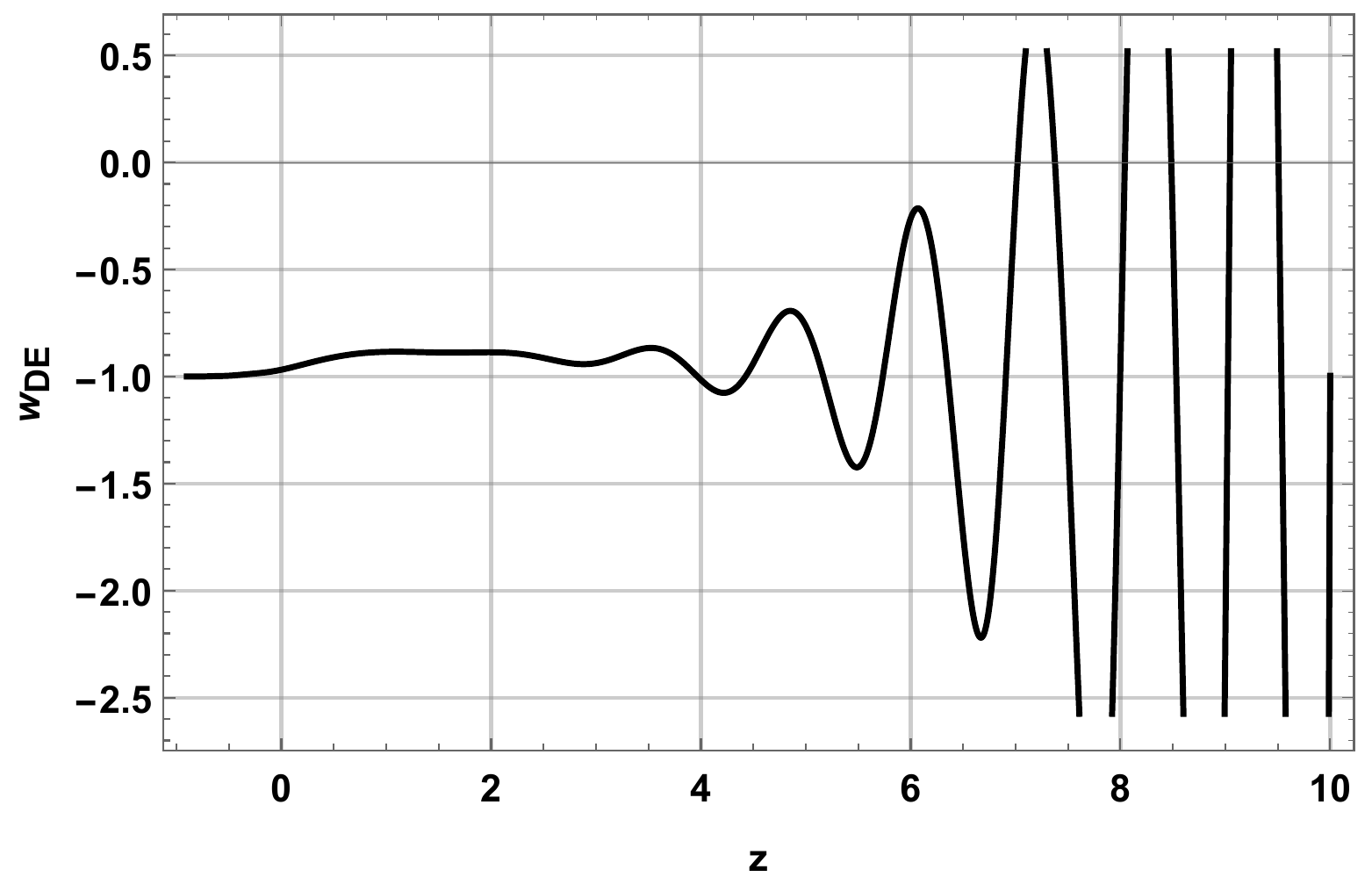}
    \caption{Evolution of the statefinder parameter $Om(z)$ (\emph{left}) and the dark energy equation of state (EoS) parameter $w_{\rm{DE}}(z)$ (\emph{right}).}
    \label{fig3}
\end{figure*}

\begin{figure*}
\centering
    \includegraphics[width=0.47\textwidth]{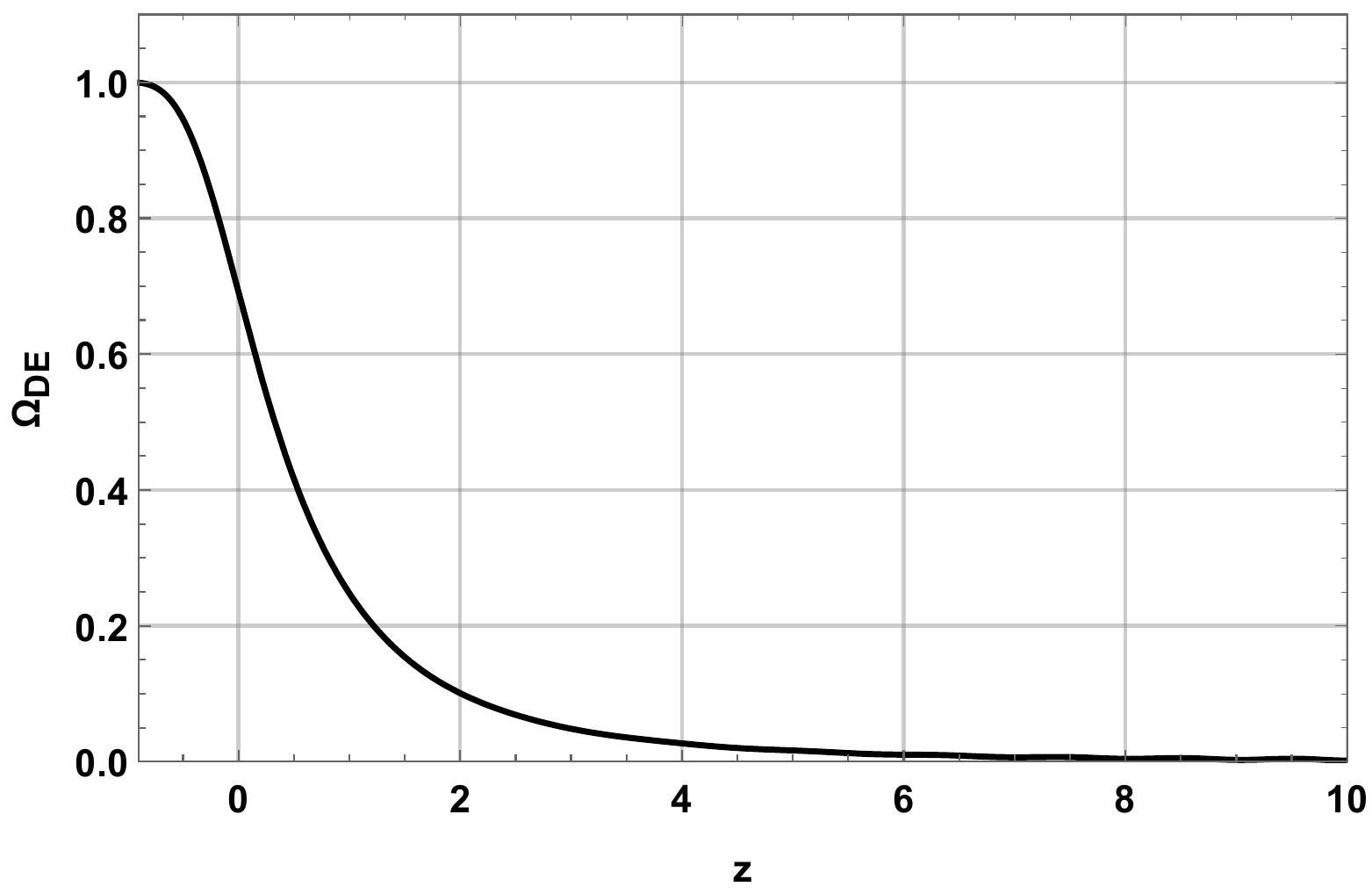}
    \includegraphics[width=0.47\textwidth]{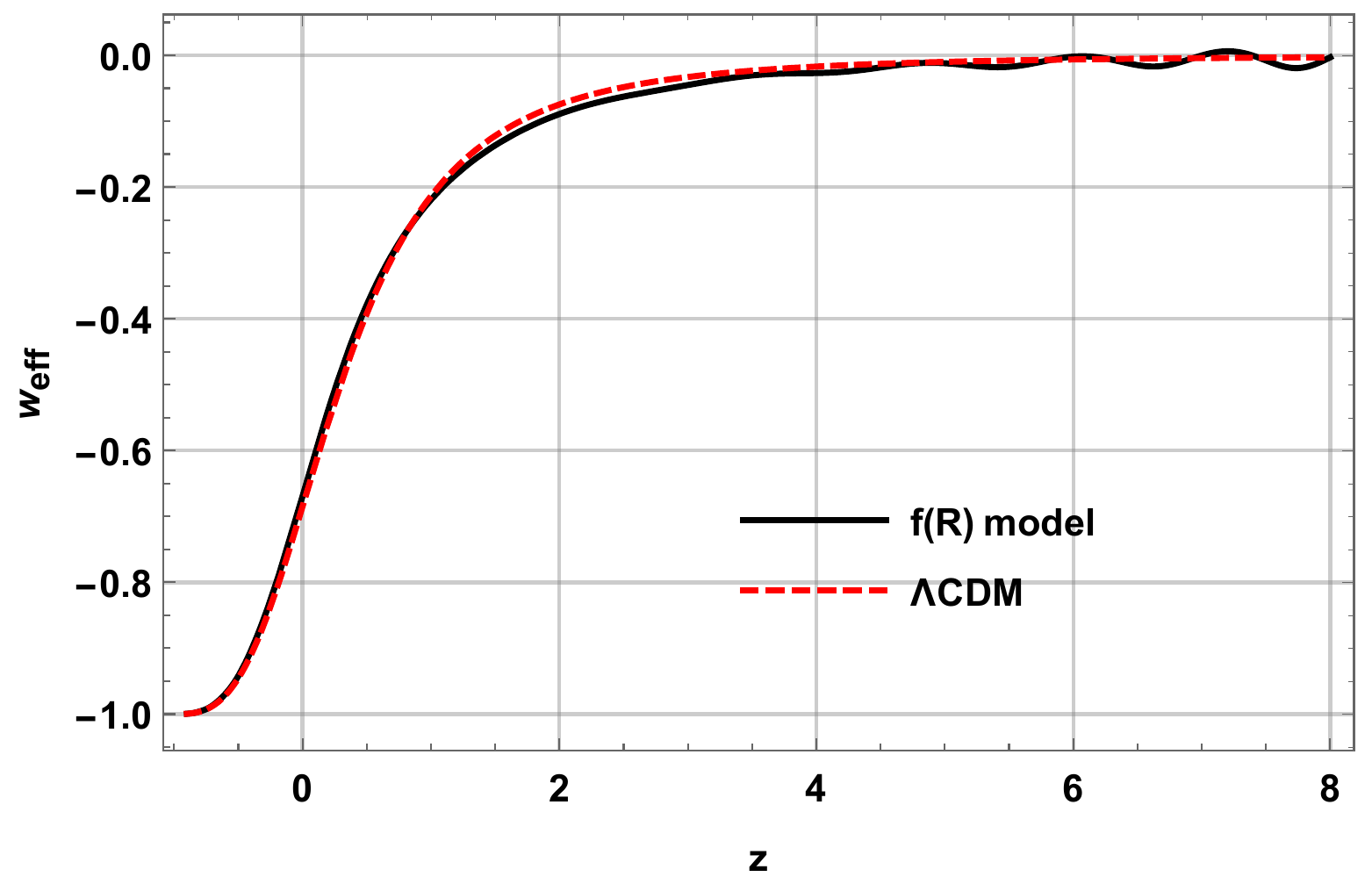}
    \caption{Change of the dark energy density parameter $\Omega_{\rm{DE}}(z)$ in terms of the redshift, $z$ (\emph{left}). Comparison between the effective EoS parameter for the exponential type $f(R)$ model (Eq.~(\ref{eq30})) and the $\Lambda$CDM model as functions of the redshift (\emph{right}).}
    \label{fig4}
\end{figure*}

\begin{figure*}
\centering
    \includegraphics[width=0.47\textwidth]{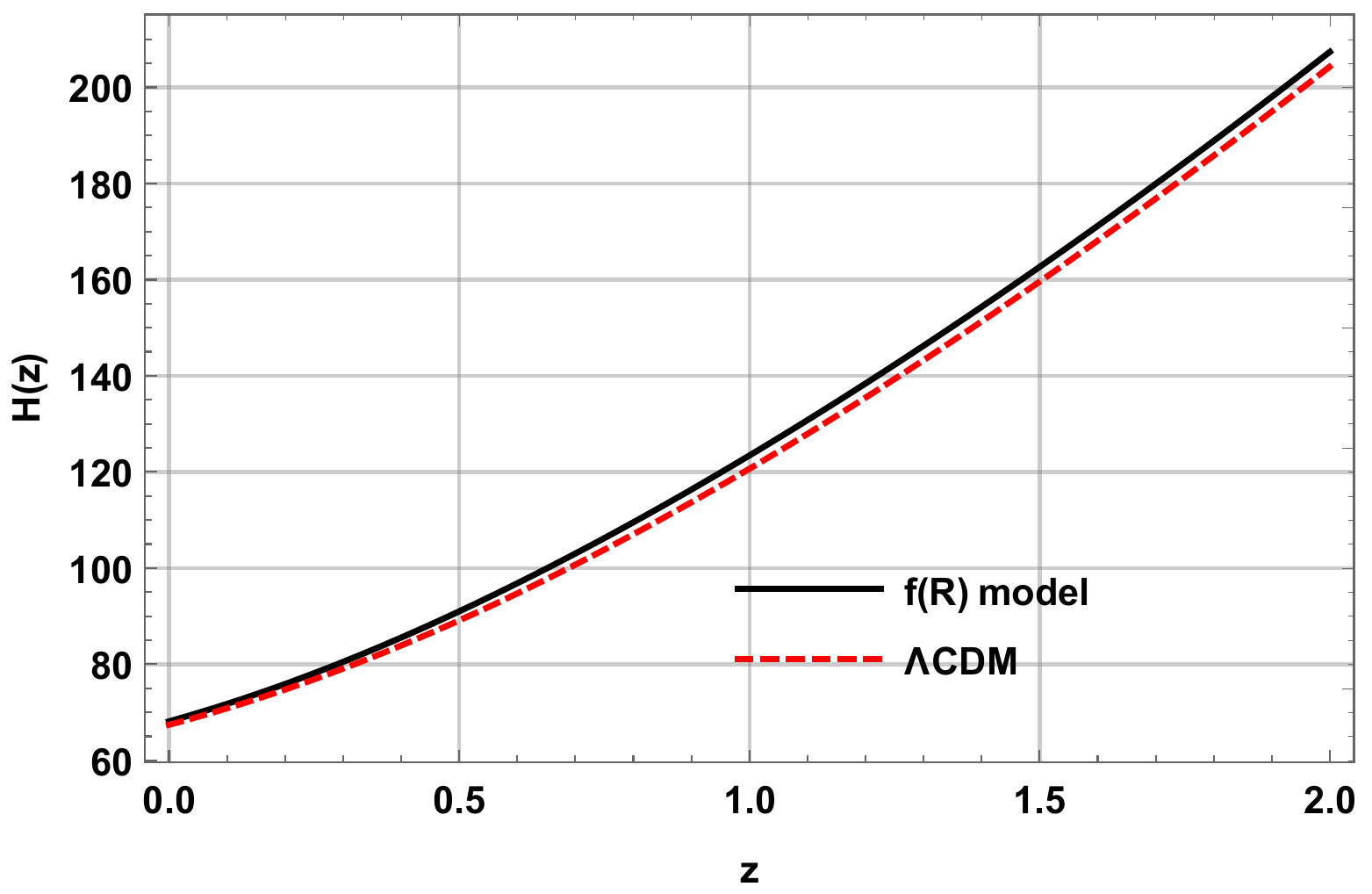}
    \includegraphics[width=0.47\textwidth]{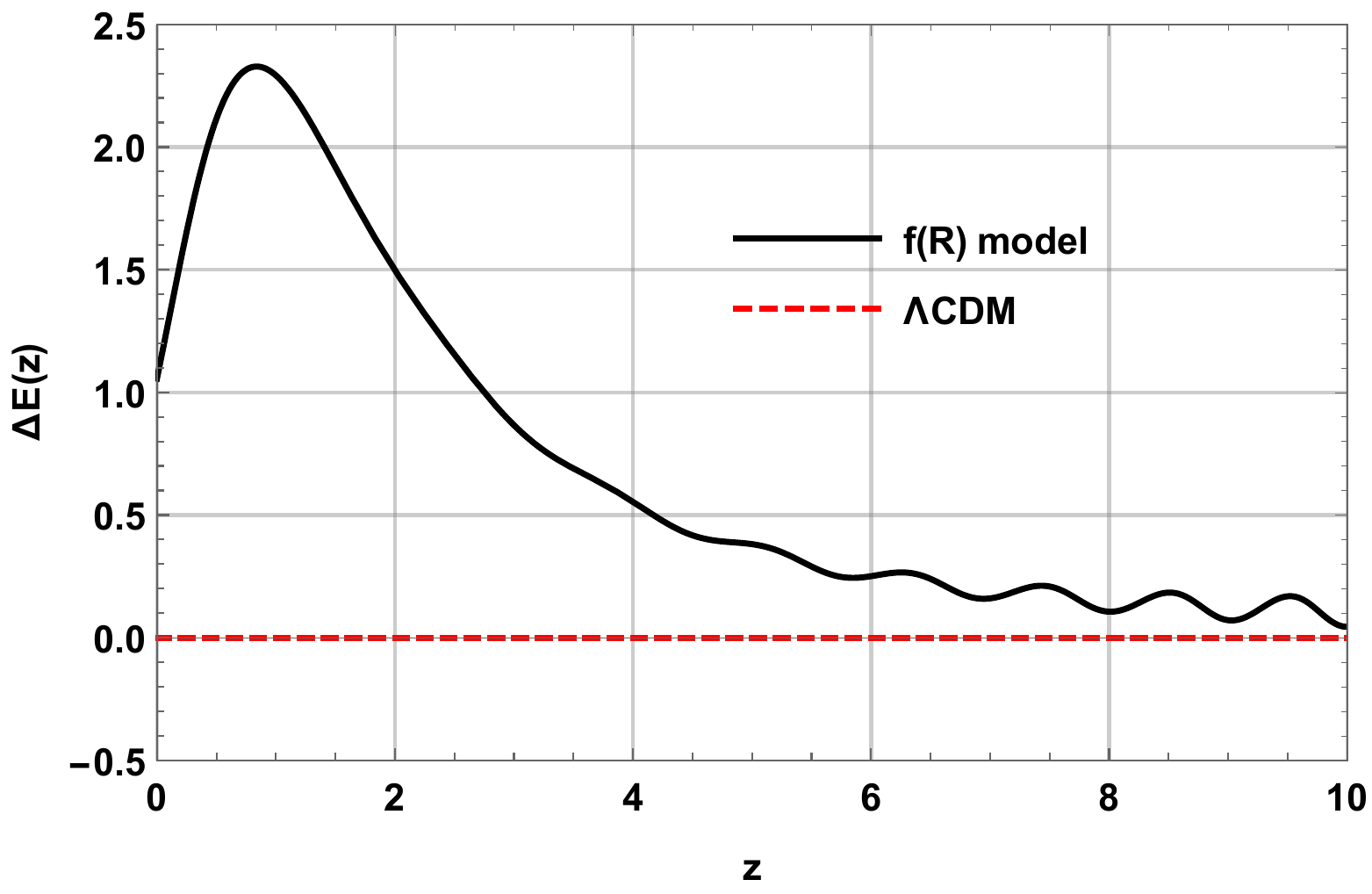}
    \caption{(\emph{Left}) Evolution of the Hubble parameter, $H(z)$, (in units of $\rm{km}\,\rm{s}^{-1}\,\rm{Mpc}^{-1}$) for the exponential type $f(R)$ model given by Eq.~(\ref{eq30}) and the $\Lambda$CDM model as functions of the redshift. (\emph{Right}) Comparison of the $f(R)$ model against $\Lambda$CDM using the dimensionless Hubble parameter through Eq.~(\ref{eq32}).}
    \label{fig5}
\end{figure*}

\begin{figure*}
\centering
    \includegraphics[width=0.47\textwidth]{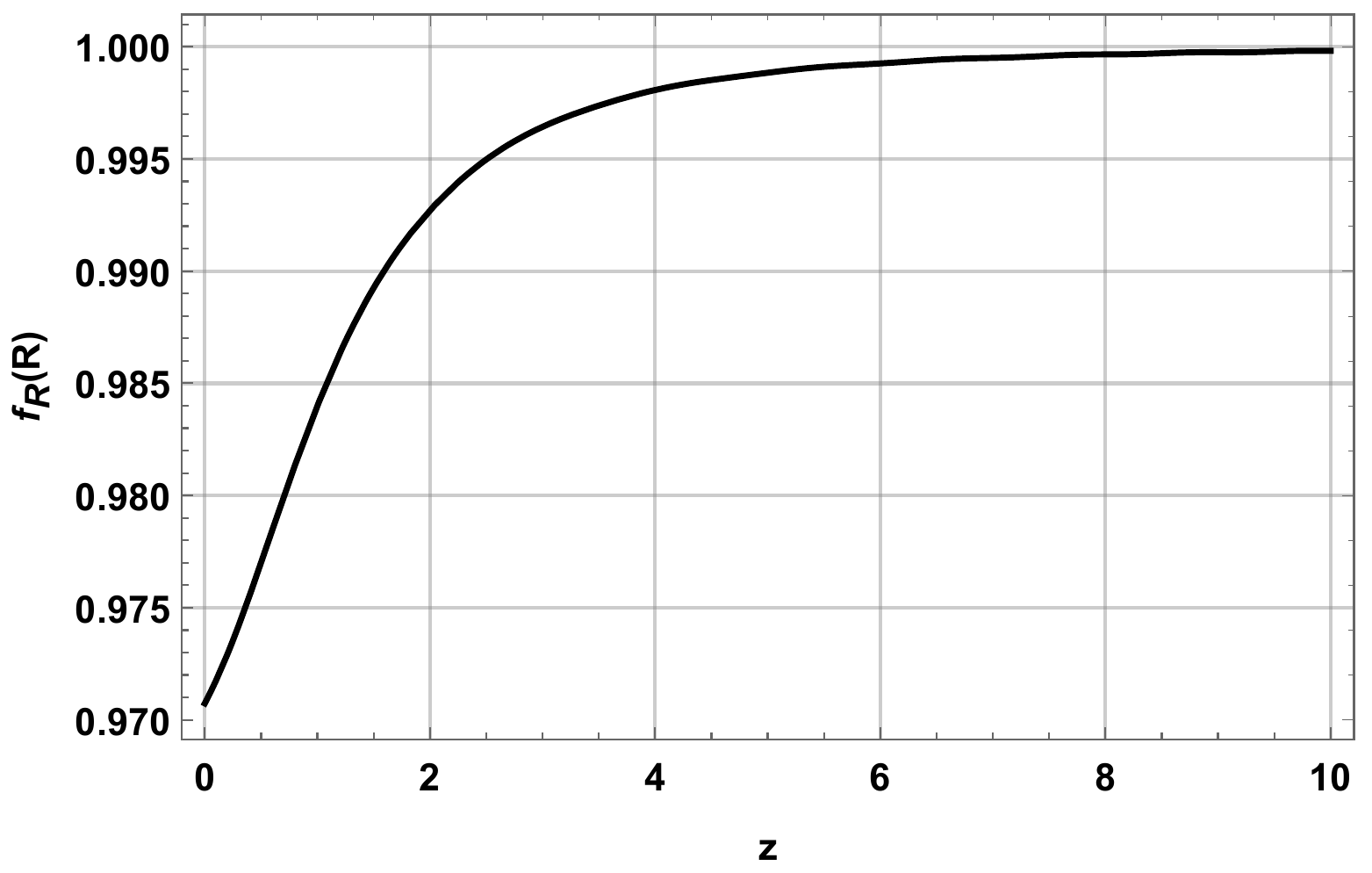}
    \includegraphics[width=0.47\textwidth]{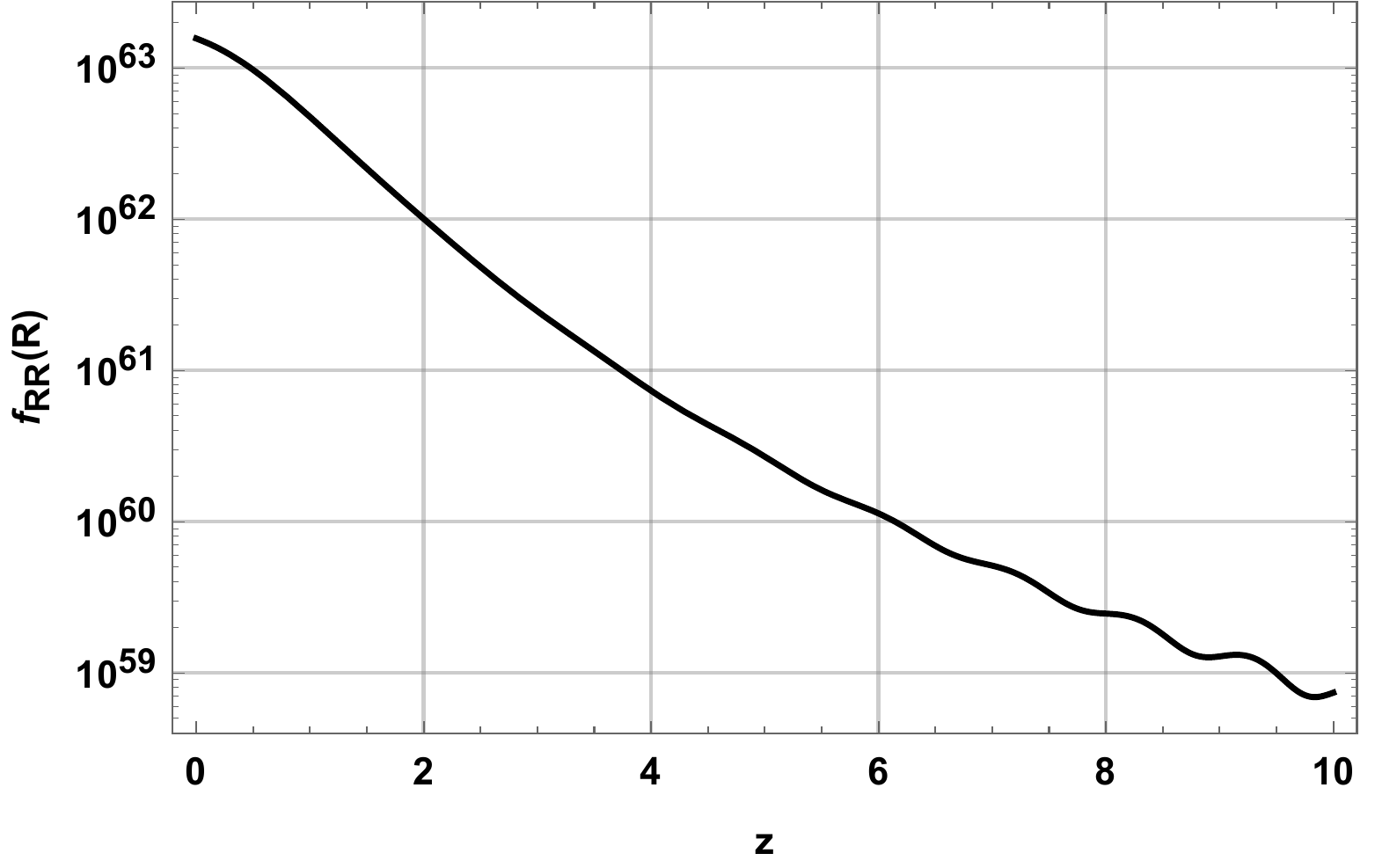}
    \caption{The terms $f_R(R)$ (\emph{left}) and $f_{RR}(R)$ (\emph{right}) as a function of the redshift.}
    \label{fig6}
\end{figure*}

\begin{figure*}
\centering
    \includegraphics[width=0.47\textwidth]{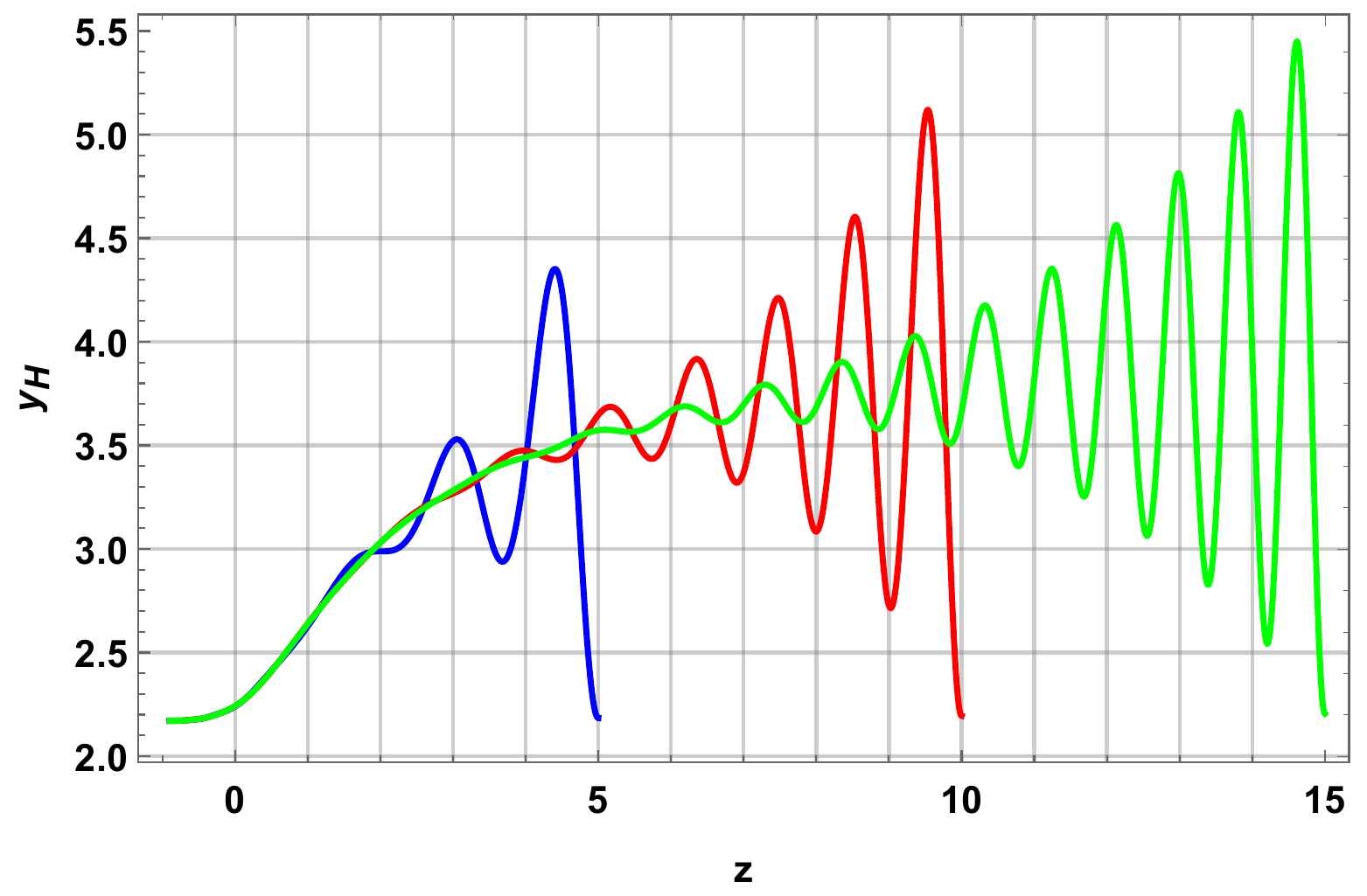}
    \includegraphics[width=0.47\textwidth]{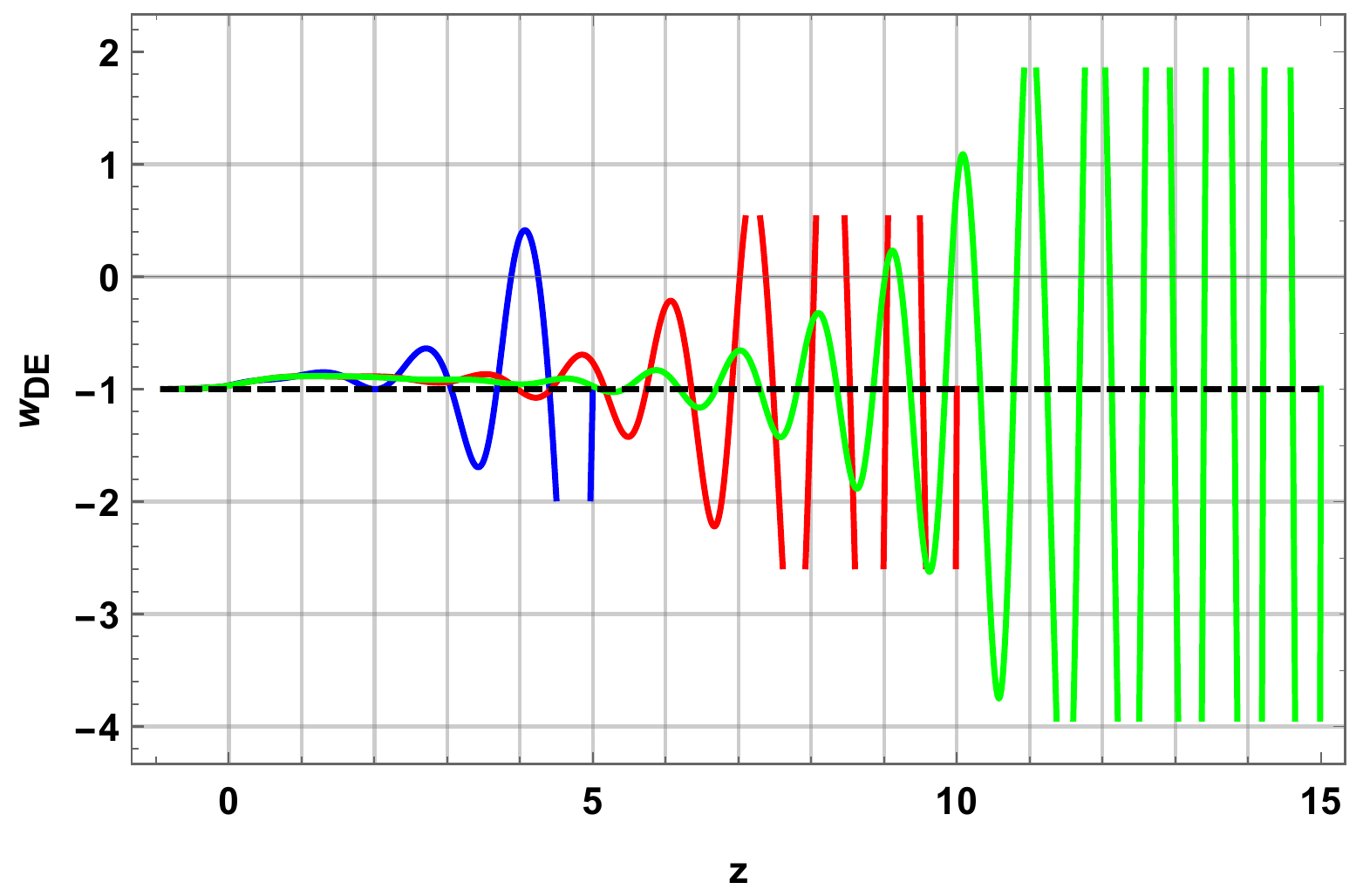}
    \caption{Changes on the evolution of $y_H(z)$ (\emph{left}) and $w_{\rm{DE}}(z)$ (\emph{right}) when the initial conditions in Eq.~\ref{eq31} are modified. In both cases we used three different choices for $z_f$: $z_f=5$ (blue), $z_f=10$ (red) and $z_f=15$ (green). The black-dashed line in the right plot, corresponds to the prediction of $\Lambda$CDM.}
    \label{fig7}
\end{figure*}

For the numerical analysis that follows, we consider well motivated physical initial conditions at $z_f = 10$, given by
\citep{bamba,odintsov6,odintsov7,oikonomou3,odintsov9}
\begin{equation}\label{eq31}
y_H(z_f)=\frac{\Lambda}{3m_s^2}\left(1+\frac{1+z_f}{1000}\right),\quad \left.\frac{dy_H(z)}{dz}\right|_{z=z_f}=\frac{1}{1000}\frac{\Lambda}{3m_s^2},
\end{equation}
where, for convenience, we choose $\Lambda\simeq 4.24\times 10^{-66}\,\rm{eV}^2$. Note that the second initial condition is always a constant. The specific numerical values are: $y_H(10)\simeq 2.1945$ and $y'_H(10)\simeq 2.17\times 10^{-3}$, respectively. In addition, to perform the numerical analysis presented in this section, it is necessary to fix the values for the model parameters. Then, following the results of Ref.~\citep{granda}, where bounds on these parameters are established, the values that we use here are: $\lambda=9.72$, $\mu^2=10^{-66}\,\rm{eV}^2$ and $\eta=0.08$; moreover $M_p=2.44\times 10^{27}\,\rm{eV}$. Taking into account the above conditions, we solve Eq.~(\ref{eq6}) for $y_H(z)$ numerically, and from Eqs.~(\ref{eq23})--(\ref{eq29}), we obtain the results displayed in Figs.~\ref{fig1}--\ref{fig7}.

First, in the left panel of Fig.~\ref{fig1} we plot the numerical solution for $y_H(z)$ and we can see that this model predicts non negligible dark energy oscillations for $z \gtrsim 4$. Further, its current value is predicted to be $y_H(0)=2.24$, which is close to the inferred value using Eq.~(\ref{eq16}), i.e., $y_H(0)=\Omega_{\rm{DE}0}/\Omega_{m0}\simeq 2.18$. The observed dark energy oscillations is more apparent in the evolution of the parameters $j$, $s$ and $w_{\rm{DE}}$, shown in Figs.~\ref{fig2} and \ref{fig3} (right panel).

Although the evolution of the parameters $q$, $Om$ and $w_{\rm{eff}}$ also exhibit these oscillations, the corresponding amplitude is smaller (see Fig.~\ref{fig1} (right panel), Fig.~\ref{fig3} (left panel) and Fig.~\ref{fig4} (right panel), respectively). In general, the amplitude of the oscillations is high at earlier epochs and decreases as the evolution proceeds.  This behavior is a generic feature of viable $f(R)$ gravity models, in particular, exponential gravity as well as a power form model (see Refs. \citep{elizalde1} and \citep{bamba} for an exhaustive analysis about this issue).

From the left plot of Fig.~\ref{fig4} we can see that the parameter $\Omega_{\rm{DE}}$ is free of oscillations. These results are similar to those found in Ref.~\citep{odintsov7} for a $f(R)$ Einstein-Gauss-Bonnet gravity model, and also in Ref.~\citep{fronimos} for a $f(R)$ Einstein-Gauss-Bonnet model with a non-minimal coupling between gravity and the kinetic term of a scalar field. However, in Ref.~\citep{odintsov6} the authors have demonstrated the absence of dark energy oscillations in the context of a $k$-Essence $f(R)$ gravity model. In the right panel of Fig.~\ref{fig4}, we have plotted the effective EoS parameter for the exponential type $f(R)$ model given by Eq.~(\ref{eq30}) and the $\Lambda$CDM model as functions of the redshift, and from it, we see that the $f(R)$ model is almost indistinguishable from $\Lambda$CDM.

Another approach to compare the $f(R)$ model studied here against the $\Lambda$CDM standard model is by using the dimensionless Hubble parameter, $E(z)=H(z)/H_0$, and calculating
\begin{equation}\label{eq32}
\Delta E(z)=100\times\left[\frac{E(z)}{E(z)_{\Lambda \text{CDM}}}-1\right],
\end{equation}
which, by definition, is zero for the $\Lambda$CDM model (red-dashed line in the right panel of Fig.~\ref{fig5}). As it is well known, $H(z)$ for the $\Lambda$CDM model is given by
\begin{equation}\label{eq33}
H(z)=H_0\sqrt{\Omega_{m0}(1+z)^3+\Omega_{DE0}+\Omega_{r0}(1+z)^4},
\end{equation}
with $H_0$ the present value of the Hubble rate, while $\Omega_{\rm{DE}0}\simeq 0.6852$, $\Omega_{m0}\simeq 0.3147$ and $\Omega_{r0}\sim 10^{-4}$.
Looking at the evolution of $\Delta E(z)$ in Fig.~\ref{fig5} (right panel), we can see that the $f(R)$ model gives $\Delta E(z)>0$, indicating that $H(z)$ is larger than the $\Lambda$CDM prediction (see also the left plot of Fig.~\ref{fig5}). The largest deviation from $\Lambda$CDM occurs around $z\simeq 0.84$, being $\Delta E(z\simeq 0.84)\simeq 2.3\%$. For $z\simeq 0$, there is a difference of the order of $1\%$. For $z>4$, the  deviation decreases in an oscillating way, approaching to the $\Lambda$CDM prediction.

In Table \ref{tab:CosmoParams} we summarize the present ($z=0$) values obtained for the cosmological and statefinder parameters studied above. For comparison, we include the corresponding values predicted by the $\Lambda$CDM model and the measurements available for some of the parameters.

\begin{table*}[h]
\centering
\caption{Current values obtained for the cosmological and
statefinder parameters using fixed and fitted (third column, BF = Best Fit) values for the free parameters of the $f(R)$ gravity model.  The values for the $\Lambda$CDM model are obtained using Eq.~(\ref{eq33}) in Eqs.~(\ref{eq25})-(\ref{eq29}). The units for $H_0$ are $\rm{km}\,\rm{s}^{-1}\,\rm{Mpc}^{-1}$. }
{\begin{tabular}{@{}ccccc@{}} \hline
Parameter & $f(R)$& $f_{\rm{BF}}(R)$ & Planck 2018 or SNe Ia & $\Lambda$CDM\\ \hline
$q_0$               & -0.504 & -0.538 & $-0.53^{+0.17}_{-0.13}$ (SNe Ia) & -0.528\\
$j_0$               & 1.011  & 1.006  & -                & 1.0002\\
$s_0$               &-0.004  &-0.002  & -                & 0.002\\
$Om_0$              & 0.330  & 0.308  & -                & 0.315\\
$w_{\rm{DE}0}$      &-0.968  &-0.999  & $-1.03\pm 0.03$    & -1 \\
$\Omega_{\rm{DE}0}$ & 0.691 & 0.6926 & $0.6847\pm 0.0073$ & 0.6852\\ 
$w_{\rm{eff}0}$     &-0.684  &-0.692  & -                  & -0.685\\
$H_0$               & 68.11 & 68.24 & $67.4\pm 0.5$      & 67.4\\ \hline
    \end{tabular}
    \label{tab:CosmoParams}}
\end{table*}

An additional important test is evaluated by looking at $f_R(R)$ and $f_{RR}(R)$ as a function of the redshift. This is shown in Fig.~\ref{fig6} from which we can see that the considered $f(R)$ model satisfies the well known viability criteria, i.e., $f_R>0$ and $f_{RR}>0$, also $f_R\rightarrow 1$ and $f_{RR}\rightarrow 0$ at $R\rightarrow \infty$.

In general, the initial conditions used for the numerical calculations described above were obtained from Eq.~(\ref{eq31}) at $z_f=10$. However, if $z_f$ is changed (keeping the same values for the free parameters), the late-time evolution of the cosmological and statefinder parameters change as well. This means that the divergence from the $\Lambda$CDM value depends strongly on the choice of $z_f$. In order to illustrate the above, in Fig.~\ref{fig7} we show the behavior for $y_H(z)$ and $w_{\rm{DE}}(z)$ using three different choices for $z_f$. It is clear that the discrepancy among the three curves at low redshift ($z<1$) is small, and in this case, the current values obtained for the cosmological and statefinder parameters are similar to those illustrated in Table \ref{tab:CosmoParams}. However, these values and the other plots are omitted by simplicity. Let us point out that in Ref.~\citep{ghosh}, the authors perform a detailed analysis about this issue using the Hu-Sawicki model.

\section{Parameter constraints from $H(z)$ observational data}\label{sec_fit}
Looking for properly characterizing the model studied here, we compare its predictions for the Hubble parameter, $H_{\rm{th}}(z)$, against a set of observational data. Let us stress that for the fit presented in this section, we have considered $H_0$ as measured by Planck \citep{planck}, which is known to be in tension with the value estimated through the local probes \citep{Riess:2021jrx} (see Ref.~\citep{DiValentino:2021izs} for a comprehensive review of this matter)\footnote{To avoid the tension, in Ref.~\citep{basilakos}, the authors  introduce a particular statistical procedure in which the chi-square estimator is not affected by the value of the Hubble constant.}. 

The statistical analysis presented in this section is performed by means of the commonly used $\chi^2$ function 
\begin{equation}\label{eq_chi2Function}
\chi^2(z;\bm{\theta}) = \sum_{i=1}^{N_d} \frac{\left(H_{\rm{th}}(z_i;\bm{\theta}) - H_{\rm{obs}}(z_i)\right)^2}{\sigma_i^2},
\end{equation}
where $N_d = 40$ is the number of observational data considered here (taken from Ref.~\citep{Cao:2021uda}), $\bm{\theta} = (\lambda,\mu,\eta)$ is the vector of free parameters of the model (see Eq.~(\ref{eq30})), and $\left( H_{\rm{obs}}(z_i),\sigma_i\right)$ corresponds to the measured value of the Hubble parameter at $z_i$ and its associated observational uncertainty.

As mentioned earlier and developed in the preceding sections, given the complexity of the $f(R)$ function, the equations were solved numerically. Furthermore, it is noted that $H_{\rm{th}}(z;\bm{\theta})$ is extremely sensible to changes of the parameters of the model, in particular to variations of $\lambda$ and $\eta$. As a consequence, we find that constraining these parameters is troublesome.

\begin{figure*}
\centering
    \includegraphics[width=0.9\textwidth]{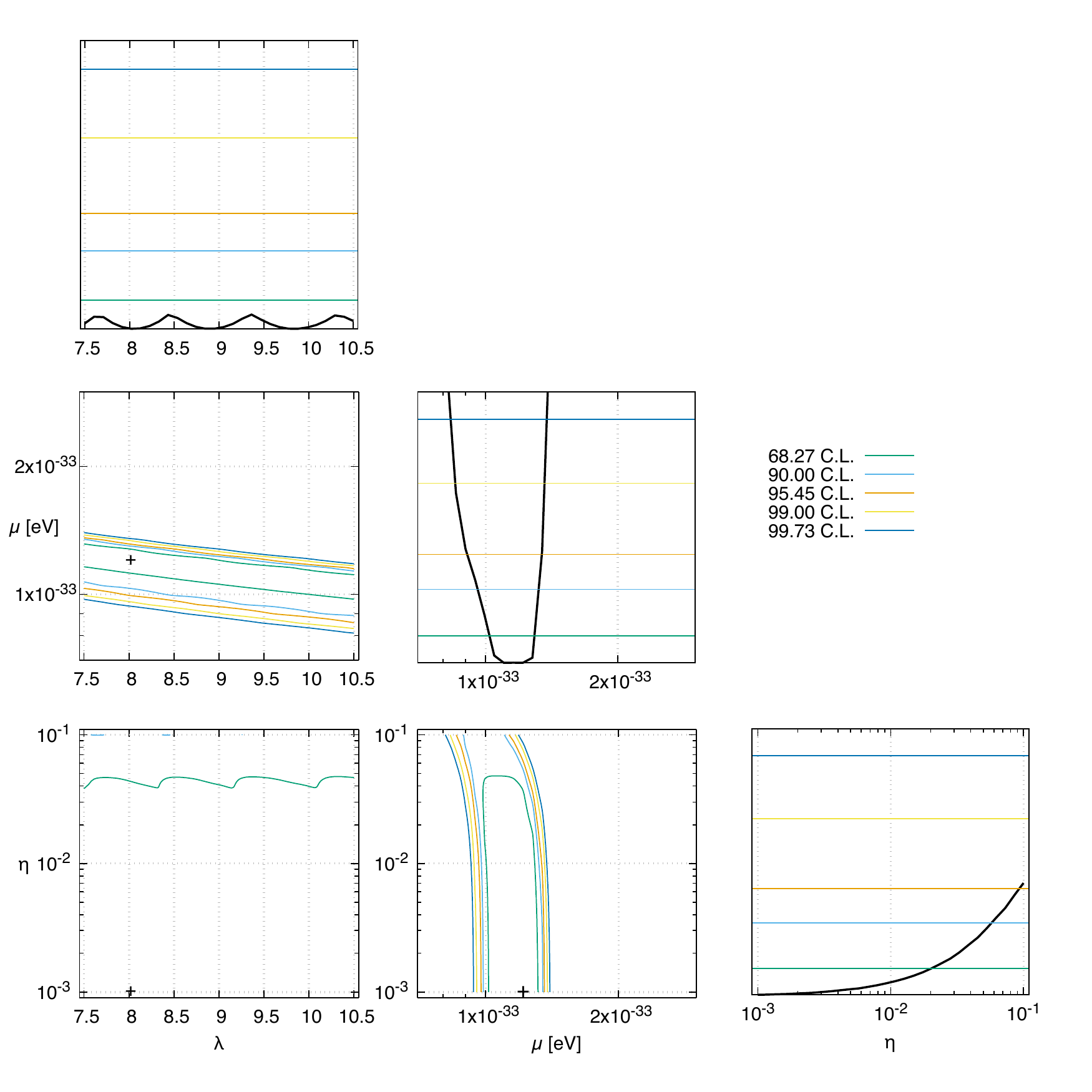}
    \caption{Allowed regions and intervals for the model parameters obtained from the $\chi^2$ analysis, at 68.27\%, 90\%, 95.45\%, 99\%, and 99.73\% C.L. The `$+$' sign in the 2D-plots locates the corresponding best fit.}
    \label{fig8}
\end{figure*}

The result of this analysis if presented in Fig.~\ref{fig8}, where an array of 2D- and 1D-plots are shown. The figures should be interpreted as follows: each 2D-plot shows the allowed regions (contours) for the displayed pair of parameters at a certain confidence level (C.L.), according to the line color. On the other hand, the 1D-plots show the marginalized allowed interval for each individual parameter, with the horizontal (colored) lines indicating the C.L.

From these plots it is evident that we obtain a well defined allowed interval for the parameter $\mu$ up to a 99.73\% C.L. and larger, while for $\eta$ there is only an upper limit: $\eta \lesssim 9\times 10^{-2}$ at 95.45\% C.L. In addition, as a confirmation of what we explained above, $\lambda$ is not well constrained by the set of observational data considered here. In particular, the uppermost 1D-plot demonstrates that, since the black line does not intersect any of the horizontal lines, every value of $\lambda$ would be equally valid to fit the data. In fact, an exploration of a larger interval than the one exhibited in the plot shows a similar low-valued oscillatory behavior with a number of local minima, preventing us to set bounds on this parameter. Clarifying this issue requires a deeper statistical analysis, considering additional observational data, and such an approach would be adrressed in a future work. In spite of this, a global minimum of the $\chi^2$ function in Eq.~(\ref{eq_chi2Function}) is found, and the best fit values obtained from the analysis, together with the allowed intervals at three particular levels of confidence are displayed in Table \ref{tab:BestFit}.

\begin{table*}[h]
\centering
\caption{Best fit and the corresponding allowed intervals for the parameters of the model. Note that the values of $\mu$ should be multiplied by $10^{-33}\,\rm{eV}$.}
\begin{tabular}{@{}ccccc@{}} \hline
Parameter & Best fit & 68.27\% & 95.45\% & 99.73\% \\ \hline
$\lambda$ & 8.02 & - & - & - \\
$\eta$    & 0.001 & $<0.02$ & $<0.09$ & - \\ 
$\mu$ & 1.22 & (1.01, 1.30) & (0.90, 1.34) & (0.81, 1.36)\\
\hline
    \end{tabular}
    \label{tab:BestFit}
\end{table*}

\begin{figure*}
\centering
    \includegraphics[width=0.7\textwidth]{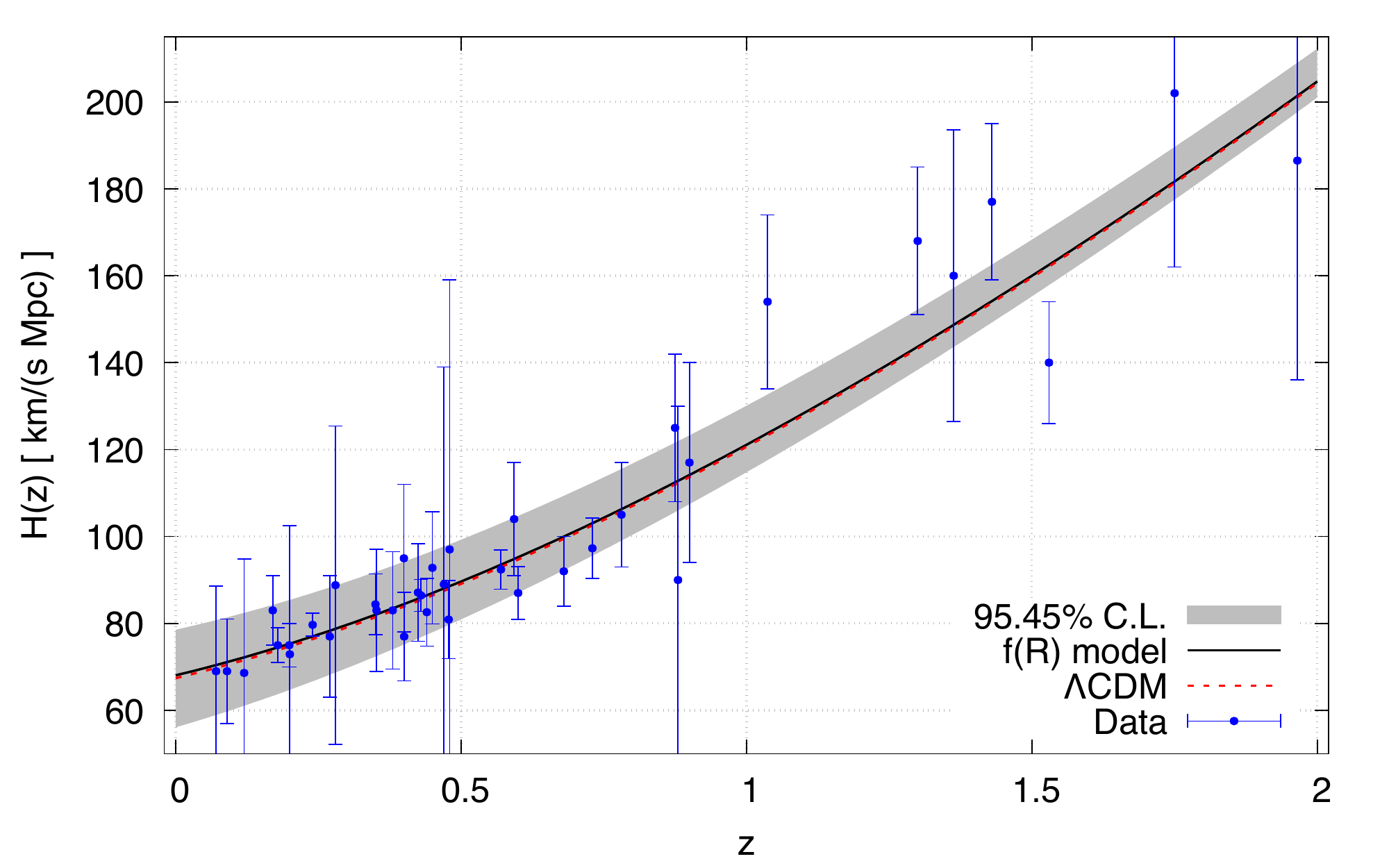}
    \caption{Best-fit $f(R)$ model prediction for $H(z)$ (black-solid line) compared against observational data (blue points --with vertical lines indicating the uncertainty--) and the $\Lambda$CDM predictions (red-dashed line). The gray-shadowed area shows the $f(R)$ model prediction when $\mu$ and $\eta$ are varied inside their 95.45\% C.L. allowed intervals.}
    \label{fig9}
\end{figure*}

The specific values of the parameters that make our model fits better the observational data (Table \ref{tab:BestFit}) can be used to compute the evolution of the Hubble parameter against the redshift for $f(R)$ in Eq.~(\ref{eq30}). This is shown in Fig.~\ref{fig9} (black-solid line), where the data are also included for comparison. This figure allows us to see that the $f(R)$ model under consideration barely marginally deviates from the $\Lambda$CDM predictions (red-dashed line), as it is usually required. In fact, when these values are used to compute the cosmological parameters and the statefinder quantities studied in the previous section at $z=0$, we obtain the numbers recorded in the third column ($f_{\rm{BF}}(R)$) of Table \ref{tab:CosmoParams}, clearly close to those predicted by $\Lambda$CDM and (for the existing cases) to the measured data. Plus, setting the parameters ($\mu$ and $\eta$) at the 95.45\% C.L.~limit values (keeping $\lambda$ at its BF), we obtain the gray-shadowed band around the best fit which, as expected, covers the vast majority of the observational data (up to their uncertainties). We have also confirmed that the best fit values of the parameters in Table \ref{tab:BestFit} satisfy the stability conditions as discussed in \citep{granda}.

In the context of this kind of analysis, in Ref.~\cite{sultana}, the authors inspect a number of well-known $f(R)$ cosmological models and constrain the corresponding parameters, namely the deviation parameter $b$ and the cosmological parameters $\Omega_m$ and $h$. The authors found that for the Hu-Sawicki and Starobinsky models, the optimized value of the deviation parameter is much closer to zero than what has been reported in the literature. In both cases, the zero value for $b$ falls safely within the $1\sigma$ (equivalent to 68.27\%) C.L., which indicates that these models coincide observationally with the predictions of $\Lambda$CDM (see Figs. 3(a) and 3(b) of \citep{sultana}). This behavior is similar to that obtained above with $\eta=10^{-3}$, since the limiting case of $\Lambda$CDM can be reached not only at high curvature but also at $\eta\rightarrow 0$ with cosmological constant $\Lambda\rightarrow e^{-1}\lambda\mu^2$. But our constraints imply too small values of $m(r)$ at current or late times ($m \ll 10^{-6}$), making it very difficult to detect measurable differences with the $\Lambda$CDM model \cite{granda}.

It is worth noting that, with this particular set of the model parameters (our best fit, Table \ref{tab:BestFit}), the evolution with $z$ of the other cosmological and statefinder quantities analysed in the previous section does not change considerably and their corresponding plot remains essentially unaffected (and the conclusions are unaffected), so we do not include new plots, for simplicity.

\section{Conclusions}\label{conclusions}
In the last years,  modified gravity theories have been established as alternative explanation to the observed late-time acceleration of the Universe, instead of the dark energy paradigm. In this context, among the wide plethora of modified gravity theories, $f(R)$ gravity emerge as one of the most studied in the literature. Following this idea, in this work we have studied the late-time evolution of the Universe for a particular viable $f(R)$ gravity model built from an exponential function of the scalar curvature. Inspired by the literature, the field equations were written in terms of a suited statefinder function, $y_H(z)$, and implementing well motivated physical initial conditions, the resulting equations were solved numerically. Also, the cosmological parameters $w_{\rm{DE}}$, $w_{\rm{eff}}$, $\Omega_{\rm{DE}}$ and $H(z)$ and the statefinder quantities $q$, $j$, $s$ and $Om(z)$ were explicitly expressed in terms of $y_H(z)$ and its derivatives (see Eqs.~(\ref{eq23}) and (\ref{eq29})). 

Furthermore, setting an appropriate set of values for the model parameters, the cosmological parameters as well as the statefinder quantities were plotted, and from them, we achieved the following results: for $y_H(z)$ we can see that this model predicts non negligible dark energy oscillations for $z \geq 4$ (see Fig.~\ref{fig1} (left panel)). Further, its current value is $y_H(0)=2.24$, which is very close to that inferred using Eq.~(\ref{eq16}), i.e., $y_H(0)=\Omega_{\rm{DE}0}/\Omega_{m0}\simeq 2.175$. The observed dark energy oscillations is more appreciable in the evolution of the parameters $j$, $s$ and $w_{\rm{DE}}$ (see Figs. \ref{fig2} and \ref{fig3} (right panel)). 

Although the parameters $q$, $Om$ and $w_{\rm{eff}}$ also present these oscillations, those are with less amplitude (Fig.~\ref{fig1} (right panel), Fig.~\ref{fig3} (left panel) and Fig.~\ref{fig4} (right panel)). In general, the amplitude of the oscillations is high at earlier epochs and decreases as the evolution proceeds. For the parameter $\Omega_{\rm{DE}}$, is clear that it is free of oscillations (see left panel of Fig.~\ref{fig4}). From the plot of $w_{\rm{eff}}$ for the given $f(R)$ gravity model and the $\Lambda$CDM model as functions of the redshift in the right panel of Fig.~\ref{fig4}, we saw that the $f(R)$ model is almost indistinguishable from $\Lambda$CDM.

In general, the cosmological and statefinder parameters at $z=0$, are shown to be compatible with Planck 2018 observations and the $\Lambda$CDM model values as exhibited in Table \ref{tab:CosmoParams}. In addition, we have compared the $f(R)$ model studied here against the $\Lambda$CDM standard model using the dimensionless Hubble parameter $E(z) = H(z)/H_0$ and calculating
$\Delta E(z)$ though Eq.~(\ref{eq32}), from which it is observed that the $f(R)$ model gives $\Delta E(z)>0$, indicating that $H(z)$ is larger than the $\Lambda$CDM prediction (see Fig.~\ref{fig5} (left panel)). The biggest deviation from $\Lambda$CDM takes place around $z\simeq 0.84$, being $\Delta E(z\simeq 0.84)\simeq 2.3\%$. For $z\simeq 0$, there is a difference of the order of $1\%$. For $z>4$, the deviation decreases in an oscillating way, approaching to the $\Lambda$CDM prediction, as depicted in Fig.~\ref{fig5} (right panel).

Additionally, $f_R(R)$ and $f_{RR}(R)$ were plotted as a function of the redshift, and it was seen that the given $f(R)$ model satisfy the well known viability criteria, ($f_R>0$ and $f_{RR}>0$), also $f_R\rightarrow 1$ and $f_{RR}\rightarrow 0$ at $R\rightarrow \infty$ (see Fig.~\ref{fig6}). We have also explored the numerical calculations using other initial conditions, but keeping the same values for the free parameters, and we observed that the late-time evolution for the cosmological and statefinder parameters change,  making evident that the deviation from the $\Lambda$CDM value strongly depends on the choice of $z_f$ (see Fig.~\ref{fig7}). However, the deviation among the curves at low redshifts ($z<1$) is small, and in this case, the current values obtained for the cosmological and statefinder parameters are similar to those illustrated in Table \ref{tab:CosmoParams}.

Finally, through familiar $\chi^2$ statistical analysis, we compared the predictions of the selected $f(R)$ gravity model for the Hubble parameter, $H(z)$, with a number of observational data. If was found that, because of the complexity of the differential equations, as well as of the $f(R)$ model itself, it is not easy to find a global minimum of the $\chi^2$ function in Eq.~(\ref{eq_chi2Function}) and that the dataset used in the analysis is not enough to unequivocally constraint the whole set of free parameters. As presented in Fig.~\ref{fig8} and Table \ref{tab:BestFit}, a well-defined allowed region is met for $\mu$ at a C.L.~larger than 99\%, while only an upper limit is obtained for $\eta$ and no bound is found for $\lambda$. In spite of this, when the best fit values are used to compute the cosmological and statefinder parameters at present time ($z=0$), the results are remarkably close to those predicted by the $\Lambda$CDM model (see Table \ref{tab:CosmoParams}). Likewise, the $f(R)$--predicted evolution of the Hubble parameter as a function of $z$ is only marginally distinguishable from that of $\Lambda$CDM (Fig.~\ref{fig9}), though this imply too small values of $m(r)$ at current or late times ($m\ll 10^{-6}$), making it very difficult to detect measurable differences with the $\Lambda$CDM model. 

The $f(R)$ gravity model considered here predicts some of the cosmological parameters to have an ill-oscillatory behavior which is an open problem for this type of models, among others. Yet, as a result of the analyses presented in this work, we can confidently say that the model is consistent with the analyzed observations, but it would be necessary to use a more complete set of observational data to obtain a best fit on the parameters of the model, e.g., using together with the measurements from the dynamics of the expansion of the Universe, $H(z)$, the growth rate of cosmic structures, $[f\sigma_8](z)$, or data from Supernovae (SNIa) Pantheon sample, among others, but that kind of analysis is beyond the scope of this work and could be addressed later.

\section*{Acknowledgments}
A.~O.~is supported by  Patrimonio Aut\'onomo-Fondo Nacional de Financiamiento para la Ciencia, la Tecnolog\'ia y la Innovaci\'on Francisco Jos\'e de Caldas (MINCIENCIAS-COLOMBIA) Grant No. 110685269447 RC-80740-465-2020, projects 69723 and 69553.



\end{document}